\documentclass[aps,prl,a4paper,twocolumn,footinbib,showpacs]{revtex4}

\usepackage{amssymb}
\usepackage{amsmath}
\usepackage{amsbsy}
\usepackage{graphicx}

\newcommand{\dr}{\textrm{d}}

\begin{document}
\title{The cosine law at the atomic scale: Toward realistic simulations of Knudsen diffusion}

\author{Franck Celestini}
\email{Franck.Celestini@unice.fr}
\affiliation{Laboratoire de Physique de la Mati\`ere Condens\'ee, UMR 6622, CNRS \& Universit\'e de Nice-Sophia Antipolis, Parc Valrose,  06108 Nice Cedex 2, France}
\author{Fabrice Mortessagne}
\email{Fabrice.Mortessagne@unice.fr}
\affiliation{Laboratoire de Physique de la Mati\`ere Condens\'ee, UMR 6622, CNRS \& Universit\'e de Nice-Sophia Antipolis, Parc Valrose,  06108 Nice Cedex 2, France}

\begin{abstract}
We propose to revisit the diffusion of atoms in the Knudsen regime in terms of a complex dynamical reflection process. By means of molecular dynamics simulation we emphasize the asymptotic nature of the cosine law of reflection at the atomic scale, and carefully analyze the resulting strong correlations in the reflection events. A dynamical interpretation of the accomodation  coefficient  associated to the slip at the wall interface is also proposed. Finally, we show that the first two moments of the stochastic process of reflection non uniformly depend on the incident angle.
\end{abstract}


\maketitle 
A century ago, Knudsen proposed a model describing the diffusion of dilute gases through cylindrical pores \cite{knudsen}. The mean free path of the molecules being larger than the pore size, the transport properties are essentially driven by the collisions with the wall. In the Knudsen approach, based on the kinetic theory of gases, all rebounds were assumed to be governed by the diffusive Lambert's cosine law of reflection \cite{knubook}. Later, the model was improved by Smoluchowski who proposed, using Maxwell's theory, that only a fraction $f$ of rebounds were diffuse, while the remaining $1-f$ were specular \cite{smolu}. The physical situations addressed by these authors gain renewed interest due to  the  emerging applications in gas separation \cite{MRS} and catalysis by means of new porous materials \cite{Davis}. In order to achieve efficient processes a thorough understanding of the mechanisms involved in transport of gas through porous membrane is required. This problem has essentially been addressed by two distinct types of numerical simulations: billiard-like simulations (BLS) and molecular dynamics (MD). In the first, the gas-wall interaction is introduced through ad hoc laws of reflection, disregarding microscopic mechanisms. Therefore, BLS is more appropriate to describe the impact of the pore geometry on the transport properties at a macroscopic scale \cite{Albo, Feres, Zschiegner}. In MD microscopic interactions are properly considered, the cost being a limitation in the spatial and time scales investigated by the simulations \cite{Skoulidas, Jakob, Nicholson}.

To combine the advantages of both kinds of simulations, one could extract from MD the realistic reflection law that would be finally incorporated in BLS. A first step in this direction has recently been made by Arya \textit{et al.} \cite{Arya}. Using  MD, these authors quantified the relationship between the phenomenological coefficient $f$ and the parameters of the Lennard-Jones potential describing gas-wall interactions, namely the wall structure and the interaction energy. Nonetheless, a realistic way of introducing $f$ in BLS has not been achieved yet. 

In this letter we propose to fill the gap between MD and BLS simulations. First, we show the robustness of the cosine law by successfully testing it against MD results. Then, the velocity distribution functions are analytically derived and successfully compared to numerical results. We re-examine the cosine law of reflections emphasizing its asymptotic nature. Through a careful analysis of the correlations between consecutive reflection events, we establish a relation between $f$ and the range of the dynamical correlations. Finally, we present an analysis of the reflection law in terms of a complex stochastic process.

\begin{figure}[h]
\includegraphics[width=0.4\textwidth]{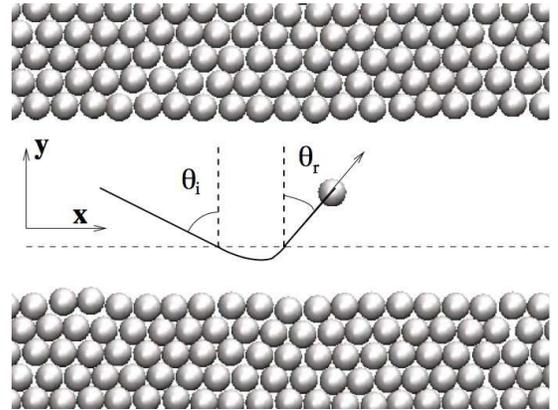}
\caption{Snapshot of the simulated system where a rebound event is schematically depicted. The dotted line shows the border between the ballistic and interacting regions.}
\label{figure1}
\end{figure}

We use MD simulations to describe the trajectory of a particle through a two-dimensional slit pore, taking into account the atomistic structure of the walls (see Fig. \ref{figure1}). The equations of motion are integrated through a standard Verlet Algorithm \cite{allen} with periodic boundary conditions  in both directions. The two different interaction potentials, between atoms  composing  the wall and between the diffusing particle and the atoms, are  of the Lennard-Jones  type: $V(r)=4\epsilon((\sigma/r)^{12}-(\sigma/r)^{6}]$. We use the shifted force potential  to  guard against energy and force discontinuities at the cut-off radius $r_c=2.5 \sigma$ \cite{allen}. For both potentials $\sigma=1$, for interactions between the wall atoms $\epsilon=1$ while $\epsilon$ takes values ranging from $0.005$ to $0.5$ for the particle-atoms potential. The mass of the atoms is fixed to $m=1$. As a consequence, lengths, energies and masses are respectively expressed in units of $\sigma$, $\epsilon$ and $m$, and times in units of  $\tau=m^{0.5} \sigma \epsilon^{0.5}$.

The walls are composed of $141$ atoms  arranged in a triangular structure (see Fig. \ref{figure1}), the  height of the pore is fixed to $6 \sigma$. Since the potential is of finite range, we can easily  define the ballistic region, in the central part of the pore, where the diffusing particle does not interact with the walls.  To characterize the ``rebounds" we therefore record the particle positions and velocities at the borders of this region (see Fig. \ref{figure1}). The Boltzmann constant is put to unity and the temperature, expressed in units of energy, is fixed to $T=0.15$, a value for which the solid phase of the wall is stable. Once the system is in thermal equilibrium we perform simulations  in the microcanonical ensemble. A typical run consists in  $3\times 10^9$ time steps  ($\Delta t = 10^{-3}$) during which we roughly compute from $10^5$ to $4\times 10^5$ rebounds  on the walls depending  on the value of $\epsilon$. 

\begin{figure}[h]
\includegraphics[width=0.4\textwidth]{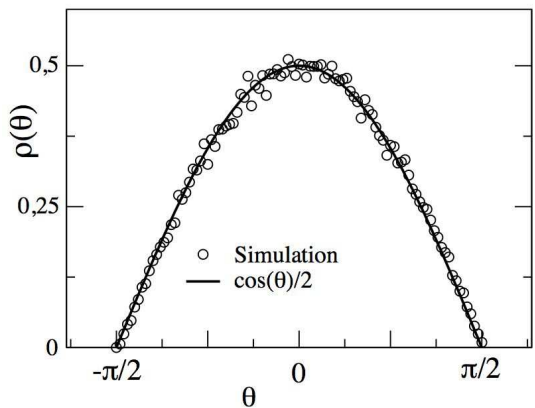}
\caption{Probability density function of the rebound angle. Simulation data (circles) obtained for  $\epsilon=0.1$ are in complete agreement with the expected cosine law (\ref{cosine}) (full line). The same agreement is observed whatever the  value of $\epsilon$.}
\label{figure2}
\end{figure}

The probability density function of the reflection angles $\rho(\theta)$ obtained numerically through an averaging over $2\times 10^5$ bounces and for $\epsilon=0.1$ is shown in Fig. \ref{figure2}. Our numerical results confirm the expected cosine law of reflection:
\begin{equation}
\rho(\theta)\dr\theta=\frac{1}{2}\cos\theta\dr\theta,\quad \rho(\sin\theta)\dr\sin\theta=\frac{1}{2}\dr\sin\theta
\label{cosine}
\end{equation}
The same remarkable agreement is observed for all the values  of $\epsilon$ used in our simulations,  showing the  robustness of the cosine law. It is worth noting that relation (\ref{cosine}) represents an asymptotic law (time $t\to\infty$) which does not preclude correlations between successive  angles of reflection. One can even obtain such ``ergodic" behavior with a pure specular law of reflection, for a non-trivial geometry of the surrounding walls. Indeed, in the limiting case of billiard problems, which constitute paradigmatic examples of Hamiltonian chaos, one generically obtains the cosine law of reflection whereas incident and reflected angles are specularly  related  \cite{Fab}. The role of short-time correlations between angles of reflection, which cannot be neglected, will be addressed in detail  in the last part of the paper.

We use the cosine law as the starting point to derive analytically the other distribution
functions describing the motion of the diffusing particle in the ballistic zone. As expected, our simulations show that the  velocity component parallel to the surface, $v_x=v\sin\theta$, is a Gaussian random  variable with variance given by the fixed temperature  (for simplicity, the mass  of the particle is 1):  
\begin{equation}
\label{distvx}
\rho_x(v_x)=\frac{1}{\sqrt{2\pi T}}\exp\left[-\frac{v_x^2}{2T}\right]
\end{equation}
The velocity can be described by two sets of random variables: $(v_x,v_y)$ or $(v, \sin\theta)$ whose distributions are related by
\begin{eqnarray}
\label{jacob}
\rho(v,\sin\theta)&=&\left\vert\frac{\partial(v_x,v_y)}{\partial(v,\sin\theta)}\right\vert\rho(v_x,v_y)\nonumber\\
&=&\frac{v}{\cos\theta}\,\rho(v_x,v_y)
\end{eqnarray}
We numerically checked that $v_x$ and $v_y$ (or, equivalently, $v$ and $\sin\theta$) are  independent random variables: $\rho(v_x,v_y)=\rho_x(v_x)\rho_y(v_y)$
($\rho(v,\sin\theta)=\rho_v(v)\rho_\theta(\sin\theta)$). Thus, the distribution $\rho_y$ can be obtained from the probability density of the modulus $\rho_v(v)=\int_{-1}^1\dr(\sin\theta)\,\rho(v,\sin\theta)$. From the normalization constraint $\int_0^\infty\dr v\,\rho_v=1$ one, indeed, obtains:
\begin{multline}
\label{norma}
\int\limits_{-1}^1\dr(\sin\theta)\,\int\limits_0^\infty\dr v\,\frac{v}{\cos\theta}\,\rho(v\cos\theta)\exp\left[-\frac{v^2\sin^2\theta}{2T}\right]\\
=\sqrt{2\pi T}
\end{multline}
where we used the Maxwell-Boltzmann distribution (\ref{distvx}). The solution of (\ref{norma}) is given by $\rho(v\cos\theta)=(v\cos\theta/T)\exp\left[-v^2\cos^2\theta/2T\right]$. Finally, we can assume for $v_y$ the following distribution
\begin{equation}
\label{distvy}
\rho_y(v_y)=\frac{v_y}{T}\exp\left[-\frac{v_y^2}{2T}\right]
\end{equation}
Due to the non-symmetric interactions experienced by the particle in the $y$-direction, 
the corresponding velocity distribution (\ref{distvy}) differs qualitatively from (\ref{distvx}).
Recall that (\ref{distvy}) is fully compatible with the diffuse law of reflection (\ref{cosine}). A similar expression of  (\ref{distvy}) had been phenomenologically obtained by Arya \textit{et al.} \cite{Arya}.

\begin{figure}[h]
\includegraphics[width=0.4\textwidth]{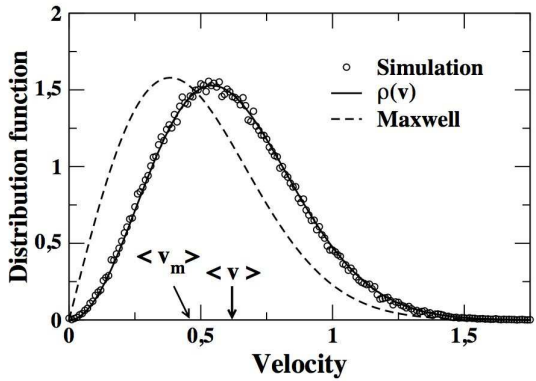}\\
\caption{Probability density function of the velocity modulus. Circles are simulation data, 
the full and dotted curves correspond respectively to $\rho_v(v)$ given in (\ref{distv}) and to  the Maxwell distribution.}
\label{figure3}
\end{figure}

Knowing $\rho_x$ and $\rho_y$, we are now able to write the explicit form of $\rho_v$:
\begin{equation}
\label{distv}
\rho_v(v)=\sqrt{\frac{2}{\pi}}\frac{v^2}{T^{3/2}}\exp\left[-\frac{v^2}{2T}\right]
\end{equation}
As expected from (\ref{distvy}), it is clear that the modulus $v$ does not obey the standard two-dimensional Maxwell distribution \footnote{For a 3D problem, we would have obtained $\rho_v(v)=(v^3/2T^{2})\exp\left[-v^2/2T\right]$}. Expression (\ref{distv}) is successfully tested against the simulation results as shown in Fig. \ref{figure3}. Here again the   agreement is quite remarkable whatever the values of $\epsilon$. For comparison the 2D Maxwell distribution is also shown in Fig. \ref{figure3}. Note that the mean value 
$\langle v\rangle  = \sqrt{8T/\pi}$ deduced from (\ref{distv}) is around 20\,\% smaller than the Maxwell mean value $\langle v_m\rangle  = \sqrt{\pi T/2}$. 
We want to draw attention to the fact that distribution (\ref{distv}) should be used in  BLS of Knudsen diffusion. Indeed, the commonly used Maxwell distribution leads to an underestimate of the diffusion constant.

\begin{figure}
\includegraphics[width=0.4\textwidth]{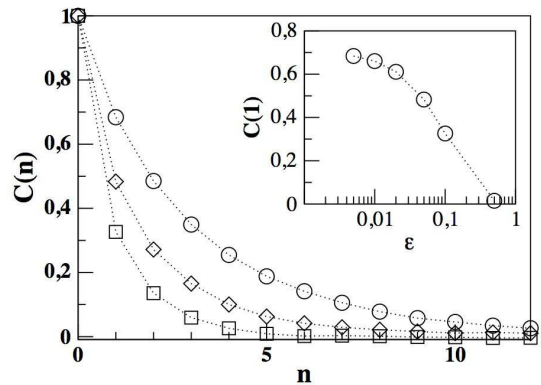}
\caption{Correlation function $C(n)$ represented for different values of $\epsilon$.
Circles, diamonds and squares correspond to $\epsilon=0.005$, $0.02$ and  $0.1$. Inset: $C(1)$ is plotted as a function of $\epsilon$ respectively.}
\label{figure4}
\end{figure}

We now turn to a thorough analysis of correlations between successive reflected angles in order to understand how the cosine law of reflection is asymptoticaly attained. We thus define the following function:
\begin{equation}
C(n) = 1 - \frac{\left\langle (\theta_p + (-1)^{n+1}\theta_{p+n})^2\right\rangle_p }{(\pi^2/2)-4}
\label{cden}
\end{equation}
$C(n)$ gives a measure  of the correlation between two angles of reflection separated by 
$n$ rebounds of the particle on the walls. By construction, $C(n)=1$ for a pure specular
regime of reflections and $C(n)=0$ in the diffusive regime. The evolution of $C(n)$ for
three values of $\epsilon$ is shown on Fig. \ref{figure4}. We observe that the
expected convergence toward the diffusive limit as $n$ increases is observed. We numerically verified that the mean resident time of the particle inside the interacting region increased with $\epsilon$. As a consequence, one can see on Fig. \ref{figure4} that the rate of convergence to $C=0$ increases with larger $\epsilon$. But it is worth noting that $C(n)$ exhibits non-zero values for a  significant number of bounces. The correlation functions exhibit an exponential behavior $C(n) = \exp(-n/n_c)$ where the characteristic number of rebounds $n_c$ gives the range of the correlation: two angles of reflection separated by a few $n_c$ rebounds are uncorrelated. For the correlation functions shown in Fig. \ref{figure4}, $n_c$ varies from $0.23$, for the largest value of $\epsilon$, to $2.92$, for the smallest. Such a large number of ``non-diffusive" (or ``quasi-specular") rebounds may drastically alter the transport process. This is  what had been phenomenologically taken into account in the Smoluchowsky model \cite{smolu} by introducing fractions $f$ and $1-f$ of atoms having diffuse and specular rebounds respectively. This fraction $f$ is called the tangential momentum accommodation coefficient and can be related to the slip coefficient at the wall \cite{maginn2,smolu}. Our approach gives an insight into the microscopic dynamics which generates the observed macroscopic behavior and provides a new interpretation of the phenomenological coefficient. In order to mimic the observed correlations, one can indeed imagine a more realistic BLS where $n_c$ rebounds are specular, after which the next rebound is diffuse and followed by another $n_c$ specular rebounds. Thus, the accommodation coefficient can be viewed as the inverse of the characteristic number of rebounds:  $f \sim 1/n_c$. 

Let us now focus on $C(1)$ which quantifies the loss of memory after a single  bounce. The inset in Fig. \ref{figure4} shows the evolution of this ``instantaneous"  correlation function with the  strength of the potential. Even for the smallest value of $\epsilon$ the reflection is never truly specular (in this case $C(1)=1$) due to the atomic structure of the wall. As expected, for increasing values of $\epsilon$ $C(1)$ vanishes, indicating a complete randomization of the reflection events. This is clearly illustrated by looking at the motion of the particle for $\epsilon=0.1$ where the absence of correlations between incident and reflected angles is evident (see EPAPS \cite{videos}, video 1). Note that the diffusive behavior stems from the fact that for almost all rebounds the particle experiences, in the interacting zone, multiple collisions with wall atoms. This is no longer the case for $\epsilon=0.01$ (see EPAPS \cite{videos}, video 2) where a unique collision event generates the rebound. 

\begin{figure}[h]
\includegraphics[width=0.4\textwidth]{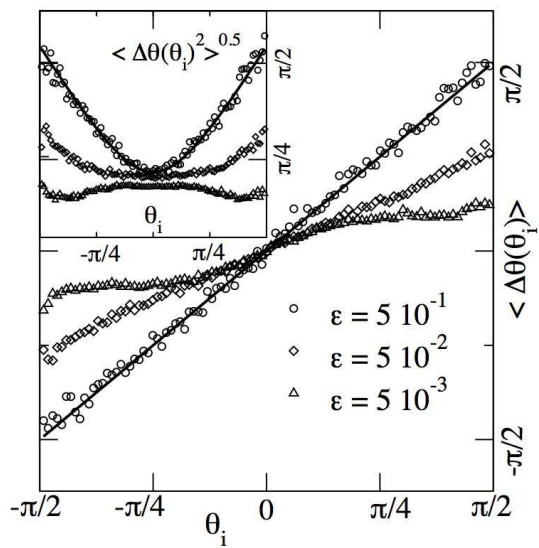}
\caption{First and second (inset) moments of the distributions $\rho_{\theta_i}(\Delta\theta)$. The full lines correspond to the purely diffusive regime. Circles, diamonds and triangles correspond respectively to $\epsilon=0.005$, $0.05$ and $0.5$.}
\label{figure5}
\end{figure}

Since the reflected angle is never completly diffusive or specular but something in between, the important quantity to know is the conditional probability
$\rho(\theta_r | \theta_i)$ to rebound with an angle $\theta_r$, the incident angle $\theta_i$ being fixed. If we want to focus on the deviation from the specular regime, it is convenient to write $\theta_r = -\theta_i +\Delta\theta$ and therefore express the conditional probability as a set of distributions $\rho_{\theta_i}(\Delta\theta)$. We computed $\left\langle\Delta\theta(\theta_i)\right\rangle$  and $\left\langle\Delta\theta(\theta_i)^2\right\rangle$, the first and second moments of these distributions, 
 and plot it for different values of $\epsilon$ in Fig \ref{figure5}. In the case of a
purely diffusive reflection we expect:
\begin{eqnarray}
\left\langle\Delta\theta(\theta_i)\right\rangle &=& \theta_i\\
\left\langle\Delta\theta(\theta_i)^2\right\rangle &=& \theta_i^2+\pi^2/4-2 
\end{eqnarray}
As shown in Fig. \ref{figure5}, the diffusive behavior is recovered for the most attractive potential ($\epsilon=0.5$) considered in this study. The opposite case of purely specular reflections, characterized by $\left\langle\Delta\theta(\theta_i)\right\rangle= 0$  and $\left\langle\Delta\theta(\theta_i)^2\right\rangle= 0$, is never reached. Indeed, as discussed above when interpreting the fact that $C(1)\neq 1$ even for $\epsilon\to 0$, the wall structure precludes the emergence of a pure specular regime. In the intermediate regimes  ($\epsilon=0.05, 0.005$), one can see that both moments non uniformly depend on the incident angle, the non-generic behavior being more important for the largest incident angles. Thus, we demonstrate here that the correlations between two successive rebounds are quantitatively sensitive to the value take by the first rebound.  We are currently developing an analytical  description of such a complex behavior in the choice of the reflected angle in an improved  BLS scheme.

In this letter, we have proposed new insight into the so-called cosine law of reflection by means of MD simulations. We confirmed its validity at the atomic scale whatever the strength of the gas-wall interaction, but only as an asymptotic law. We examined two crude ``ergodic" approximations in the studies of Knudsen diffusion: which (\textit{i}) assume Maxwell velocity distribution functions, (\textit{ii}) neglect the correlations between consecutive rebounds. We gave the correct distributions, Eq. (\ref{distvy}) and (\ref{distv}), actually involved in the Knudsen regime. By defining an appropriate correlation function (\ref{cden}), we carefully analyzed the strong correlations occurring in the dynamical process of reflection. We revealed a characteristic correlation rebound number $n_c$ that has been linked to the commonly used accommodation coefficient $f$. Finally, we performed the complete characterization of the stochastic rebound process and emphasized the non-trivial dependence of the conditional probability $\rho(\theta_r\vert\theta_i)$ on the incident angle. This work should pave the way for new studies of Knudsen transport using improved BLS schemes as proposed in this letter.

\begin{acknowledgments}
We gratefully acknowledge fruitful discussions with A. ten Bosch.  We also wish to thank G. Batrouni, A. ten Bosch and S\'ebastien Tanzilli for their careful reading of the manuscript.
\end{acknowledgments}

\end{document}